\documentclass[twocolumn, reprint, superscriptaddress,pra]{revtex4-1}
\usepackage{graphicx}
\usepackage{epstopdf}
\graphicspath{{./figures/}}
\usepackage{amssymb}
\usepackage{float}
\usepackage{physics}
\usepackage{commath} 
\usepackage{amsmath}
\usepackage[table,xcdraw]{xcolor} 
\usepackage{multirow}
\usepackage{hhline}
\usepackage{tabularx}
\usepackage{enumitem}
\newlist{arrowlist}{itemize}{3}
\setlist[arrowlist]{label=$\Rightarrow$}
\usepackage{hyperref}

\begin{document}
\title{Noise-specific beats in the higher-level Ramsey curves of a transmon qubit}
\author{L. A. Martinez}
\affiliation{Lawrence Livermore National Laboratory, 
CA, USA }
\author{Z. Peng}
\affiliation{Michigan State University, East Lansing, MI}
\author{D. Appel\"{o}}
\affiliation{Michigan State University, East Lansing, MI}
\author{D. M. Tennant}
\author{N. Anders Petersson}
\author{J. L DuBois}
\author{Y. J. Rosen}
\affiliation{Lawrence Livermore National Laboratory, 
CA, USA }

\date{Nov 11, 2022}

\begin{abstract}
 In the higher levels of superconducting transmon devices, and more generally charge sensitive devices, $T_2^*$ measurements made in the presence of low-frequency time-correlated $1/f$ charge noise and quasiparticle-induced parity flips can give an underestimation of the total dephasing time. The charge variations manifest as beating patterns observed in the overlay of several Ramsey fringe curves, and are reproduced with a phenomenological Ramsey curve model which accounts for the charge variations. $T_2^*$ dephasing times which more accurately represent the total dephasing time are obtained. The phenomenological model is compared with a Lindblad master equation model. Both models are found to be in agreement with one another and the experimental data. Finally, the phenomenological formulation enables a simple method in which the power spectral density (PSD) for the low-frequency noise can be inferred from the overlay of several Ramsey curves.
\end{abstract}

\maketitle

Owing to its exponentially suppressed charge sensitivity, the superconducting transmon qubit has established itself as a leading candidate for NISQ-era quantum computing systems \cite{arute2019quantum,kandala2017hardware}. Nevertheless, qubit dephasing and relaxation caused by interactions with various noise sources continue to limit the time available for error-free quantum gate operations. Measurements of the noise plaguing superconducting qubits have been vital in illuminating and mitigating the various sources of noise \cite{bertet2005dephasing,sung2019non,bylander2011noise,schlor2019correlating,sung2021multi,yan2012spectroscopy,serniak2019direct,yuge2011measurement,von2020two,burnett2019decoherence}. For example, photon shot noise has been established as one of the dominant sources of dephasing. This is especially important in 3D-transmon architectures \cite{wang2019cavity} where qubit coherence is dominated by pure dephasing, and mitigation through filtering and thermalization has been demonstrated \cite{sears2012photon}. Sources of noise have also been attributed to the interaction of two-level-systems (TLS) residing at the material interfaces and surfaces  \cite{burnett2019decoherence}, non-equilibrium quasi-particles \cite{riste2013millisecond,serniak2018hot}, and anomalous charge drift across the junction \cite{christensen2019anomalous}. More recently, cosmic rays were identified as a source of correlated errors suggesting new obstacles in the way of reaching fault tolerant quantum computation \cite{wilen2021correlated}.

As quantum computing platforms strive towards larger Hilbert spaces, the higher levels of superconducting transmon qubits may hold promise for quantum computing applications  \cite{wu2020high,liu2017transferring}. Admittedly, most of the success of the transmon can be attributed to its first energy transition (0-1) since the higher levels of the transmon exhibit shorter dephasing times -- suggesting charge noise magnitudes which prevent their usage for applications in quantum computing. However, characterizing dephasing in the presence of correlated charge noise can lead to a misrepresentation of the time available for quantum state manipulation. Transmons and charge sensitive qubits plagued with this problem exhibit seemingly random Ramsey curves, which form beat patterns \cite{peterer2015coherence} -- a signature of correlated low-frequency charge noise. In these devices, characterization of the dephasing time requires explicit consideration of the charge noise.

The Ramsey pulse sequence \cite{ramsey1950molecular} is at the core of qubit noise measurements as it transforms the qubit into a sensitive probe susceptible to both classical and quantum noise, and it is used to acquire Ramsey curves which provide a measurement of the total dephasing parameter $T_2^*$ \cite{vion2002manipulating}. A single Ramsey curve measurement consists of an initial $R_x(\pi/2)$ pulse which initiates the qubit into a quantum superposition state between two qubit levels, a free evolution period where its phase evolves, and a second $R_x(\pi/2)$ pulse that projects the phase onto a single qubit level for readout. During the free evolution period of a charge sensitive state, charge noise coupling to the qubit modulates its resonant frequency while energy relaxation leads to state decay; both manifest as stochastic variations in the phase of the quantum superposition state. 

If the charge fluctuations are zero-mean Gaussian distributed, averaging over $N$ individual Ramsey curves reduces the standard error by $1/\sqrt N$. This is the case when the charge noise exhibits a white noise power spectral density. In this case, the $T_2^*$ value extracted from fitting an exponentially decaying sinusoid to the average of several Ramsey curves is an accurate representation of the total dephasing time. However, in the presence of charge drift (time-correlated low-frequency $1/f^\alpha$ charge noise for $\alpha>0$) \cite{christensen2019anomalous} and quasiparticle induced parity switches \cite{riste2013millisecond}, the $T_2^*$ value extracted from fitting a decaying sinusoid to the average of multiple Ramsey curves can result in an underestimation of the total dephasing time. In this case, the canonical approach averages several Ramsey curves over slow-varying stochastic charge noise -- resulting in an effective destructive interference. This underestimation of $T_2^*$ also occurs in transmon qubits where, despite being generally regarded as charge insensitive devices, the higher levels can exhibit sufficiently large charge sensitivity to be affected \cite{tennant2022low}.

We begin this letter with a demonstration of how the charge noise manifests as a beating pattern in the higher levels of a 3D transmon system \cite{wu2020high}, followed with a quick discussion on how noise leads to ambiguous $T_2^*$ measurements. We then present a phenomenological Ramsey curve model which explicitly accounts for slow-rate charge parity flips and charge drift. We extract a corrected $T_2^*$ and compare it to the value obtained with the canonical method. Next, we use the phenomenological model to identify the noise power spectral density (PSD), from which we can determine the likely origin of the noise. Finally, the phenomenological model is compared with both experimental data and full Lindblad master equation simulations. We find that both models reproduce the experimental data.

\begin{figure}[t]
   \centering
   \includegraphics[width = .31\textwidth]{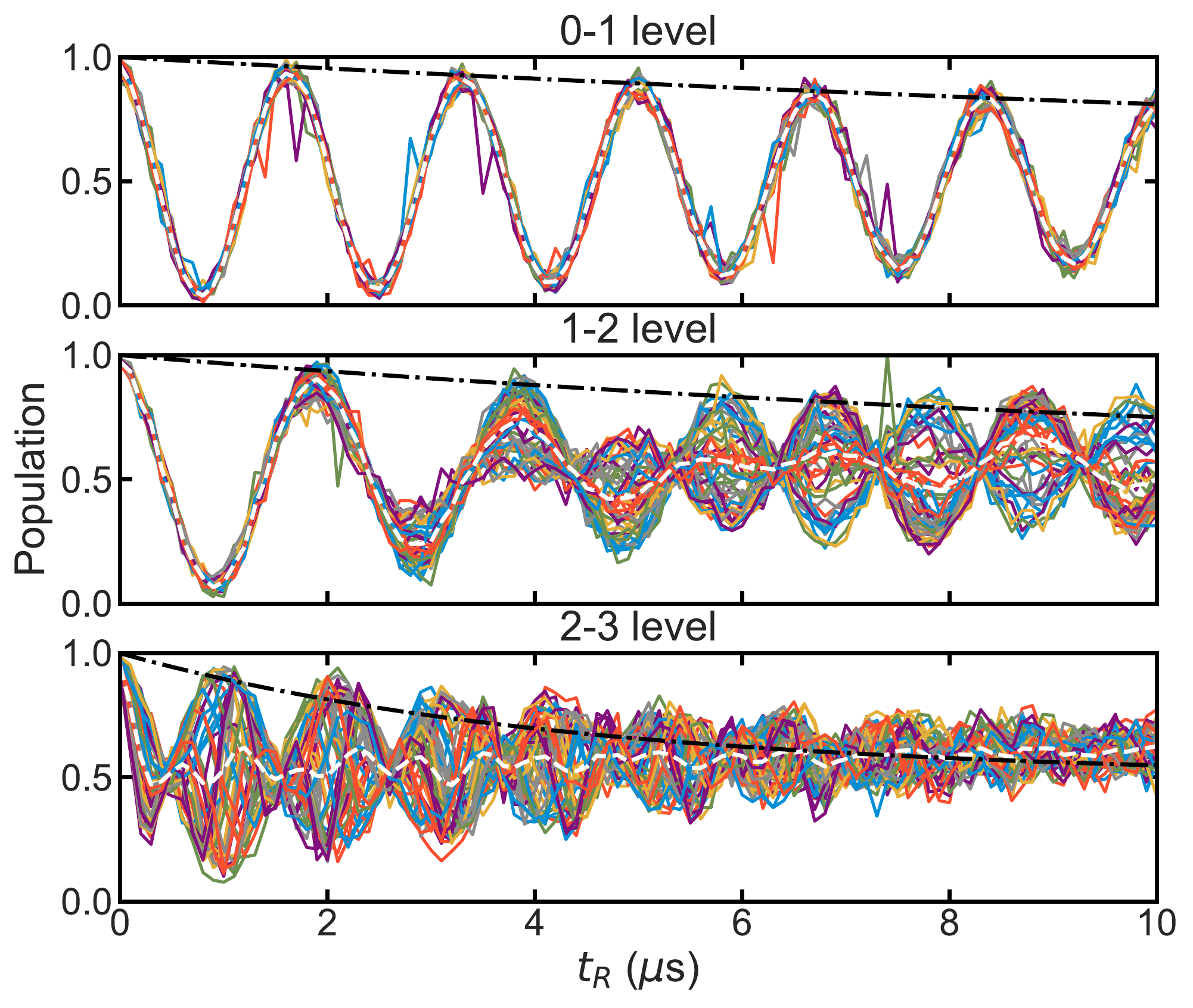} 
   \caption{Overlay, decay envelope (black dashed-dotted lines), and respective averages (white-dashed lines) of 50 Ramsey curves collected sequentially on the first 3 levels of a transmon qubit with 500 kHz Ramsey detuning. Beat patterns are observed on levels with larger dispersion values. Starting at top, the charge dispersion for the 0-1, 1-2, and 2-3 levels are 1.7, 62, and 1305 kHz, respectively. }
   \label{fig:3levels_ramseys}
\end{figure}

Quantum systems utilizing the higher levels of transmon qubits, and more generally charge sensitive devices, plagued with charge noise exhibit random Ramsey curves between repeated measurements. The overlay of 50 Ramsey curves along with their respective averages (white-dashed lines) for each of the first three levels of a transmon are shown in figure \ref{fig:3levels_ramseys}. The charge noise coupled to the qubit results in seemingly random individual Ramsey curves, which when superimposed yield a sinusoidal beating pattern for the 1-2 and 2-3 levels. The first level has a low-dispersion value calculated to be 1.7 kHz. In this case, the 50-curve overlay does not exhibit a beating pattern and averaging over more curves is beneficial. However owing to the higher dispersion value of 62 kHz measured for the 1-2 level, it is more sensitive to charge variations. The increased charge sensitivity leads to the beating pattern observed in the 50-curve overlay that is otherwise not visible on the averaged curve. This behavior is significantly more pronounced for the 2-3 level which has a measured dispersion of 1.3 MHz. Using the canonical approach of fitting a decaying sinusoidal to the averaged traces, results in ${\tilde T}_2^* = 2.7~\mu$s and ${\tilde T}_2^* = 0.3~\mu$s for the 1-2 and 2-3 levels, respectively. This approach of extracting the total dephasing time leads to values that are dominated by low-frequency noise and, therefore, underestimate the total dephasing time. 

\begin{figure}[t]
   \centering
   \includegraphics[width = .47\textwidth]{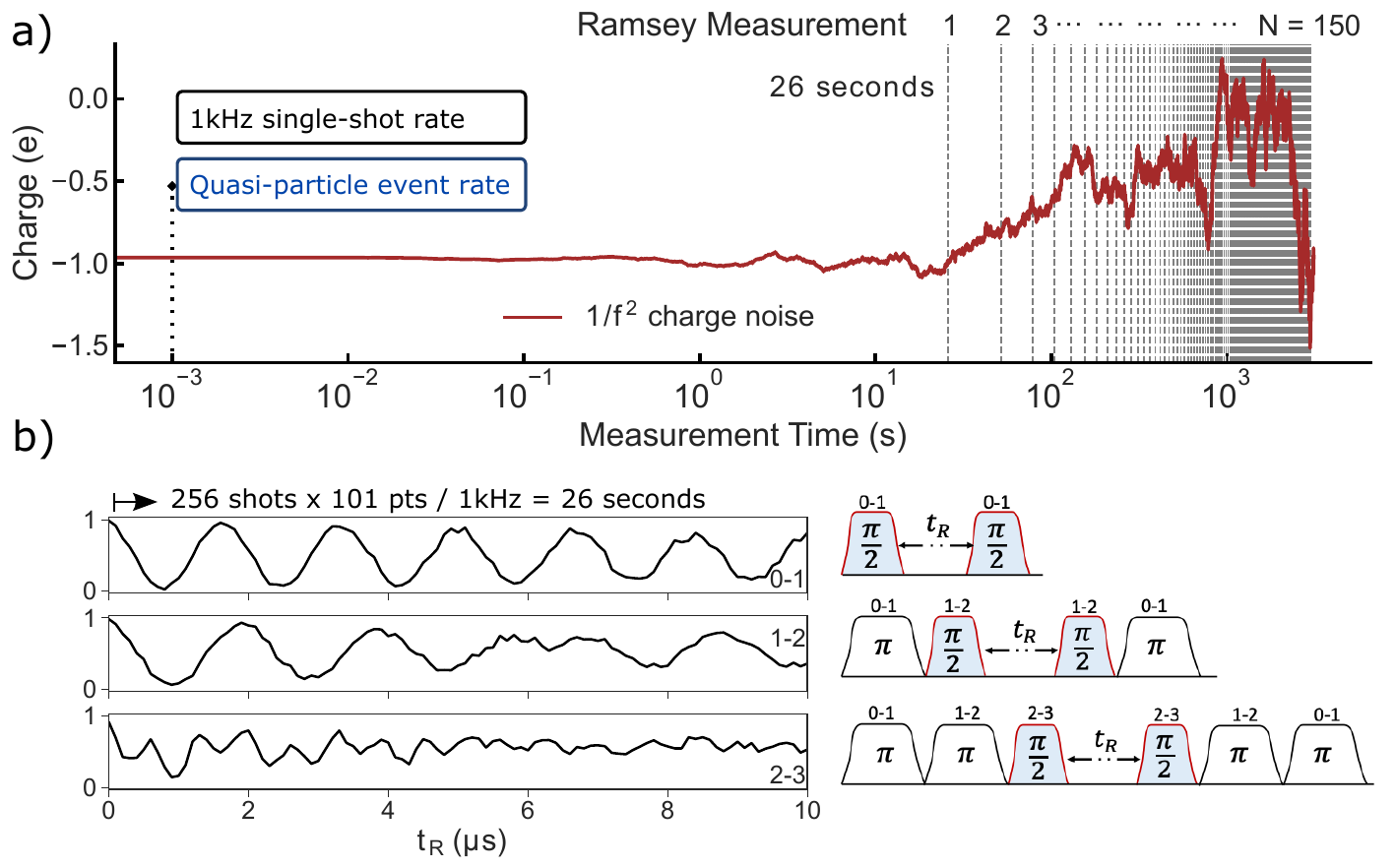} 
   \caption{ a) The relevant time scales. With a 1 kHz single-shot rate, a Ramsey curve is collected approximately every 26 seconds; $1/f^2$ (simulated) charge noise becomes significant after the first Ramsey curve measurement. b) Multi-level Ramsey measurement sequence in a transmon qubit \cite{suppmaterial} with $E_j/Ec = 50$. Each Ramsey curve consists of 101 free-evolution time values ($t_R)$ with 256 single shots collected for each value. The first Ramsey curve is collected on the 0-1 level, the second on the 1-2, the third on 2-3, and this sequence repeats until 50 Ramsey curves are collected for each level giving a total of $N = 150$ curves. Corresponding pulse sequences are shown on the right. The 1-2 and 2-3 levels use additional pre (post) $\pi$-pulse sequences to excite (de-excite) the qubit. Ground state is used for readout \cite{peterer2015coherence}.}
   \label{fig:meas_scheme}
\end{figure}

To understand how the beating patterns arise, we consider the relevant timescales involved. The Ramsey curves shown in figure \ref{fig:3levels_ramseys} were collected on the first three levels of a superconducting 3D transmon CQED system \cite{wu2020high, suppmaterial} suffering from quasiparticle poisoning \cite{riste2013millisecond} and charge drift \cite{christensen2019anomalous} (figure \ref{fig:meas_scheme}). The single-shot measurement repetition rate was set to 1 kHz, 101 free-evolution time values and 256 shots per value were collected, and a single Ramsey experiment was completed every 26 seconds. At these timescales, the quasiparticle-induced parity flips occurring at a rate of approximately $\sim 1$ kHz \cite{riste2013millisecond} lead to an effective destructive interference when averaging over multiple shots (or multiple Ramsey curves). A characteristic feature of $1/f^\alpha$ noise is that its correlation length increases with the exponent $\alpha$; $\alpha = 0$ (white noise) is uncorrelated while $\alpha =2$ (brown noise) is time-correlated. Due to this correlation, as the number of Ramsey curves included in the average increases the contribution of low-frequency $1/f^2$ charge noise on the measured $T_2^*$ becomes more significant. Furthermore, looking at an individual 2-3 level Ramsey curve (figure \ref{fig:meas_scheme}b) one might incorrectly conclude that its $T_2^*$ is too short for practical applications. Note, in practice long acquisition times arise from the lack of high fidelity single-shot readout which requires averaging over several measurements, and/or measurements that are made with slow sampling rate systems.

Next, we formulate a phenomenological model which takes into account charge-parity flips and low-frequency noise and, thereby, yields $T_2^*$ measurements which more accurately represent the time available for coherent state manipulation. Previous experimental observations have demonstrated quasiparticle events in transmon qubits which cause the instantaneous qubit frequency to switch symmetrically about the mean qubit frequency at milli-second time intervals \cite{riste2013millisecond}; $f_q(t) = \bar{f}_{q,ij}+\epsilon^{\text{max}}_{ij}\cos(2\pi n_g(t)+p(t)\pi)$ where the second term is the dispersion relation of the $ij^\text{th}$ level given by the approximation of the Mathieu solutions in the transmon limit \cite{koch2007charge,cottet2002implementation}, $\epsilon^{\text{max}}_{ij}$ is the level's maximum dispersion, $n_g$ is the charge bias across the transmon's Josephson junction, and $p \in \{0,1\}$ is the parity. Since a quasiparticle event can be assumed to change the qubit frequency instantaneously relative to the single shot rate, in the limit where the rate of parity-switching events is much slower than $1/T_2^*$, the qubit frequency will appear to switch between two dominant spectral components (i.e. charge parity bands). The averaged behavior of the two spectral components, assumed to be equally displaced in frequency from the average qubit frequency ${\bar \omega_q}$\cite{riste2013millisecond}, can be modeled in a (detuned) Ramsey curve measurement by the relation,

\begin{equation}
{\cal P}\sim\frac{1}{2} \left[\cos((\Omega_R + \epsilon_{ij}(t)) t_R)+\cos((\Omega_R-\epsilon_{ij}(t))t_R)\right],
\label{eq:paritybands}
\end{equation}
where ${\cal P}$ is the qubit population during a Ramsey measurement, $\Omega_R \equiv {\bar \omega_q} - \omega_d$ is the Ramsey detuning frequency with respect to the average qubit frequency, $\omega_d$ is the qubit drive frequency, $t_R$ is the Ramsey free-evolution time, and 
\begin{equation}
\epsilon_{ij}(t) = \epsilon^{\text{max}}_{ij}\cos(2\pi n_g(t)).
\label{eq:dispersion}
\end{equation} 
 The charge bias across the junction is subject to time-dependent stochastic charge noise $n_g(t)$. Here, $t$ represents the dynamical time variable and should not be confused with the Ramsey free-evolution time $t_R$. 
 
The total dephasing time that is of interest corresponds to the Ramsey curve decay envelope. Note, determining the appropriate functional form of the decay envelope in the presence of noise processes occuring at timescales on-par or faster than the time it takes to perform a single Ramsey experiment has been previously addressed \cite{Ithier2005decoherence}. To demonstrate how our method accounts for the low-frequency charge noise, we assume the charge noise responsible for the decaying envelope in a single Ramsey experiment is represented by a white-noise power spectral density, as is typical in practice. In this case, the uncorrelated Gaussian noise processes are incorporated with a $e^{-t/T_{2}^*}$ term \cite{martinis2003decoherence,wangsness1953dynamical}. Combining equations \eqref{eq:paritybands} and \eqref{eq:dispersion} with the exponential decay term and using the sum of cosine identity gives the expression for the phenomenological model for the qubit population during a Ramsey curve measurement,

\begin{equation}
 {\cal P}(t,t_R)  = \frac{1}{2} \left[1+e^{-t_R/T_{2}^*}\cos(\Omega_R t_R)\cos(\epsilon_{ij}(t)t_R)\right],
 \label{eq:model}
 \end{equation}
normalized to represent the excited state probability. This expression is consistent with the fact that dephasing processes can be factorized, so long the high-frequency depolarization noise is non-deterministic \cite{Ithier2005decoherence,makhlin2003dephasing}.

Equation \eqref{eq:model} forms the phenomenological model and it represents the main result of this letter. It can be used in charge sensitive systems to extract a $T_2^*$ measurement which excludes quasiparticle and charge noise contributions. Importantly, in contrast to the canonical decaying sinusoid formulation this formulation for the Ramsey state population includes an additional term, $\cos(\epsilon_{ij}(t)t_R)$, which carries an explicit time dependance. For this reason, averaging over the slow stochastic variations in $\epsilon_{ij}(t)$ will give rise to an effective destructive interference. Measurements of $T_2^*$ should include only the decaying envelope in \eqref{eq:model}, i.e., $\exp(-t/T_2^*)$ since we have assumed Gaussian noise processes. This is accomplished by fitting an exponential to the envelope of the beating pattern observed in the overlay of several Ramsey measurements. Fitting to the envelope (black dashed-dotted lines in figure \ref{fig:3levels_ramseys}), the best fit $T_2^*$ values for the 1-2 and 2-3 levels were found to be $14.5~\mu$s and $4.3~\mu$s, respectively. These values are significantly different from the values reported above, obtained with the canonical approach: ${\tilde T}_2^* = 2.7~\mu$s and ${\tilde T}_2^* = 0.3~\mu$s, respectively.

\begin{figure}[h]
   \centering
   \includegraphics[width = .4\textwidth]{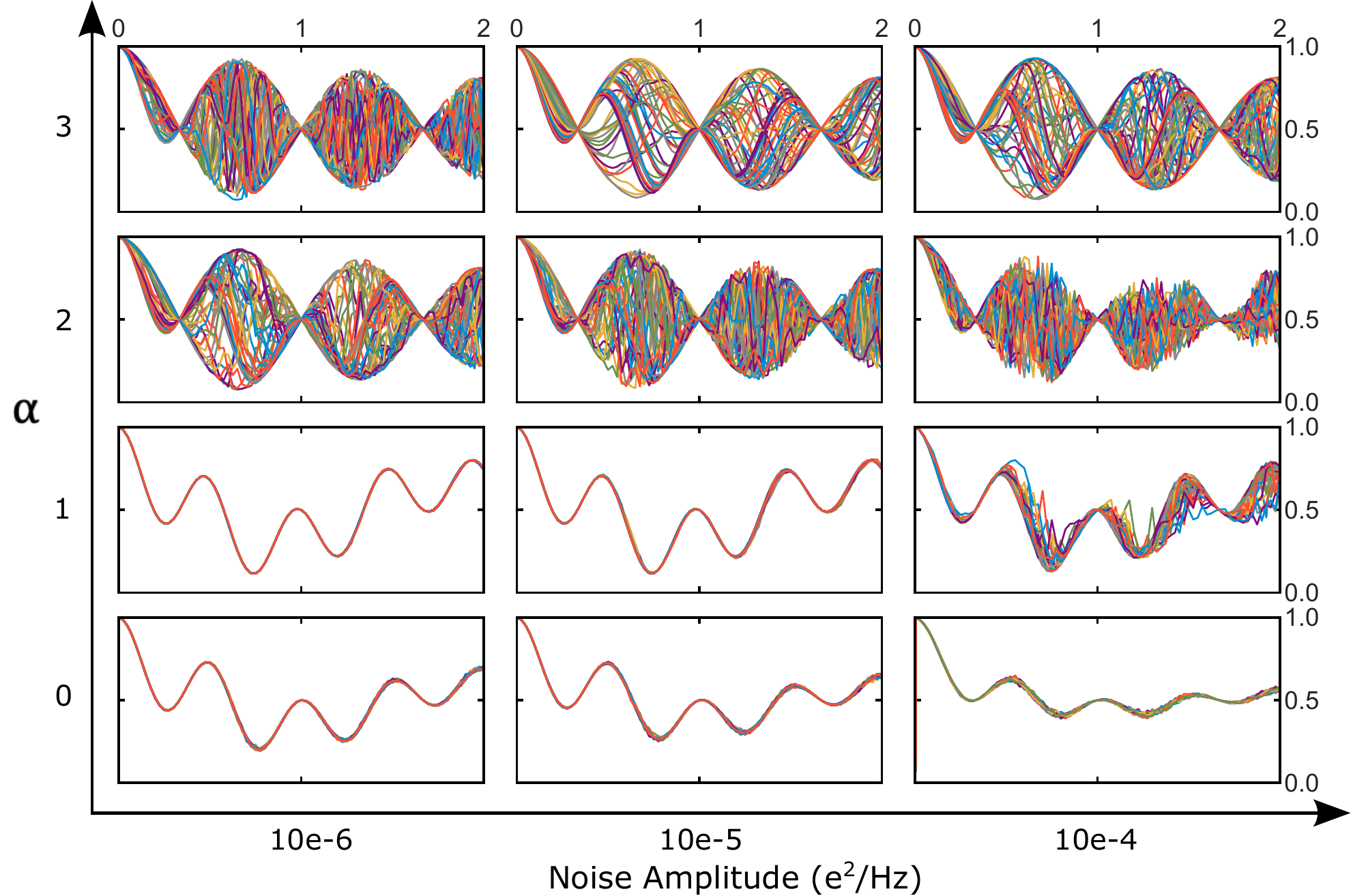} 
   \caption{Ramsey curves generated with equation \ref{eq:model} for selected values of both the noise exponent $\alpha$ and noise amplitude $S_q(1Hz) = A$. Curves shown are for the 2-3 level. Noise simulation sampling rate was 1 kHz, and $\Omega_R = 750$ kHz.}
   \label{fig:2D_parameter_sweep}
\end{figure}

We now use equation \eqref{eq:model} to explore Ramsey curve phenomenology in relation to the 1 kHz single-shot rate for selected $S(f) = A/f^\alpha$ noise models of $n_g(t)$. Figure \ref{fig:2D_parameter_sweep} shows the results of selected 2-3 level Ramsey curves generated with the simulated noise for exponents $\alpha = \{0,1,2,3\}$ \cite{timmer1995generating}, and selected noise amplitudes $A$ in the $[10^{-6},10^{-4}]e^2/\text{Hz}$ range. A key observation is highlighted. For $\alpha = 0$ and $\alpha = 1$, at low amplitudes the charge fluctuations are insufficient to cause a significant change between the individual Ramsey curves, while at higher amplitudes the larger charge variations scramble the phase of the Ramsey curves at single-shot times scales. That is, for the 1 kHz single-shot repetition rate, the short correlation time associated with $\alpha = 1$ is practically indistinguishable from uncorrelated ($\alpha = 0$) noise. Therefore, not much variability among the 50 Ramsey curves is observed - they all resemble the average value. However as the noise moves from uncorrelated towards long-time correlated noise ($\alpha =0 \rightarrow \alpha =3$), the low-frequency contribution of the $1/f^\alpha$-type charge noise becomes significant, and the low-frequency charge variations manifest with the experimentally observed beating pattern. 

\begin{figure}[h]
   \centering
   \includegraphics[width = .3\textwidth]{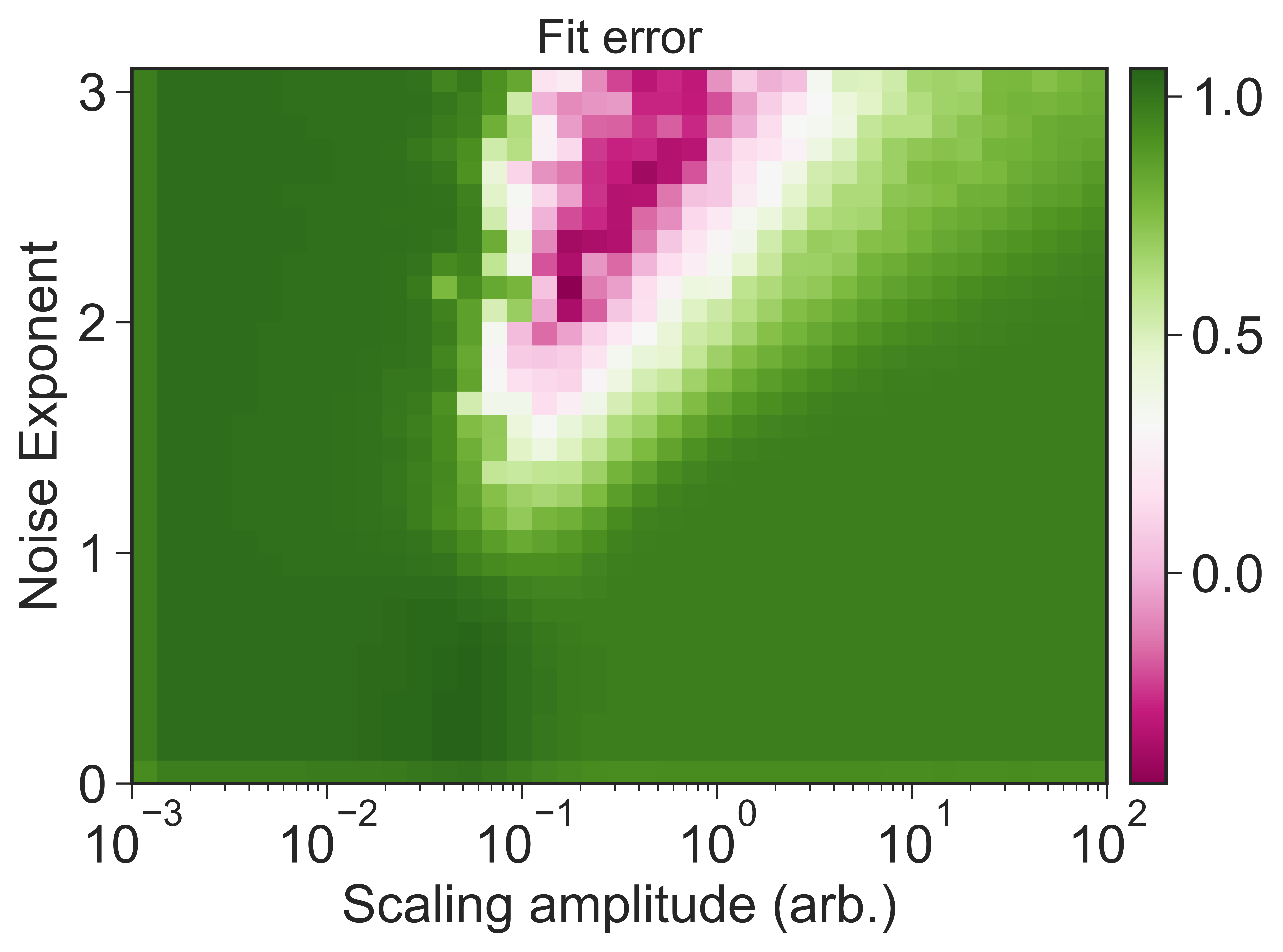} 
   \caption{Logarithm of the sum of the fit error between the simulated and experimental Ramsey curves, see text.}
   \label{fig:fft_errors}
\end{figure}

Next, we find the noise power spectral density which best reproduces the experimental data. First, simulated timeseries of $S(f) = A/f^\alpha$ were generated \cite{timmer1995generating} by performing a 2-dimensional sweep of the noise exponent ($\alpha$) and a scaling amplitude ($a$). The scaling amplitude simplifies the noise simulations over various $\alpha$. Since converting the noise timeseries into the frequency domain results in unique $A$ for each $(\alpha,a)$ pair, this approach avoids solving the inverse problem. Note, $A\propto a^2$ \cite{suppmaterial}. The sampling rate for the noise generation was set to 1 kHz, $\alpha$ varied from $0\leq\alpha\leq 3$ in steps of 0.1, and $a$ varied logarithmically from $10^{-3}$ to $10^2$. Each simulated noise timeseries, $n_g(t)$ (cyclic in units of $2e$), was then substituted into equations \eqref{eq:dispersion} and \eqref{eq:model}, and the corrected $T_2^*$ values obtained by fitting the envelopes of the beat patterns were used. The best-fit noise spectral density was determined by calculating the mean-squared difference between the experimental and simulated curves (simulated with equation \eqref{eq:model}) for three metrics: the FFT's of the individual Ramsey curves, the envelopes of 50-curve overlay, and the 50-curve averages. The FFTs of the individual Ramsey curves capture the high frequency features, the envelope of the 50-curve overlay capture the decay, and the averages capture the long-term behavior. The fit-error was defined as the equal-weighted sum of the three metrics \cite{suppmaterial}. The logarithm of the fit-error is plotted in figure \ref{fig:fft_errors}. Locating the minimum yielded the best fit $\alpha = 2.0 \pm 0.2$ and noise magnitude $S_q(1Hz) =  (2.7\pm 2.3) \times10^{-5}~e^2$/Hz \cite{suppmaterial}. 

\begin{figure}[t]
   \centering
   \includegraphics[width = .31\textwidth]{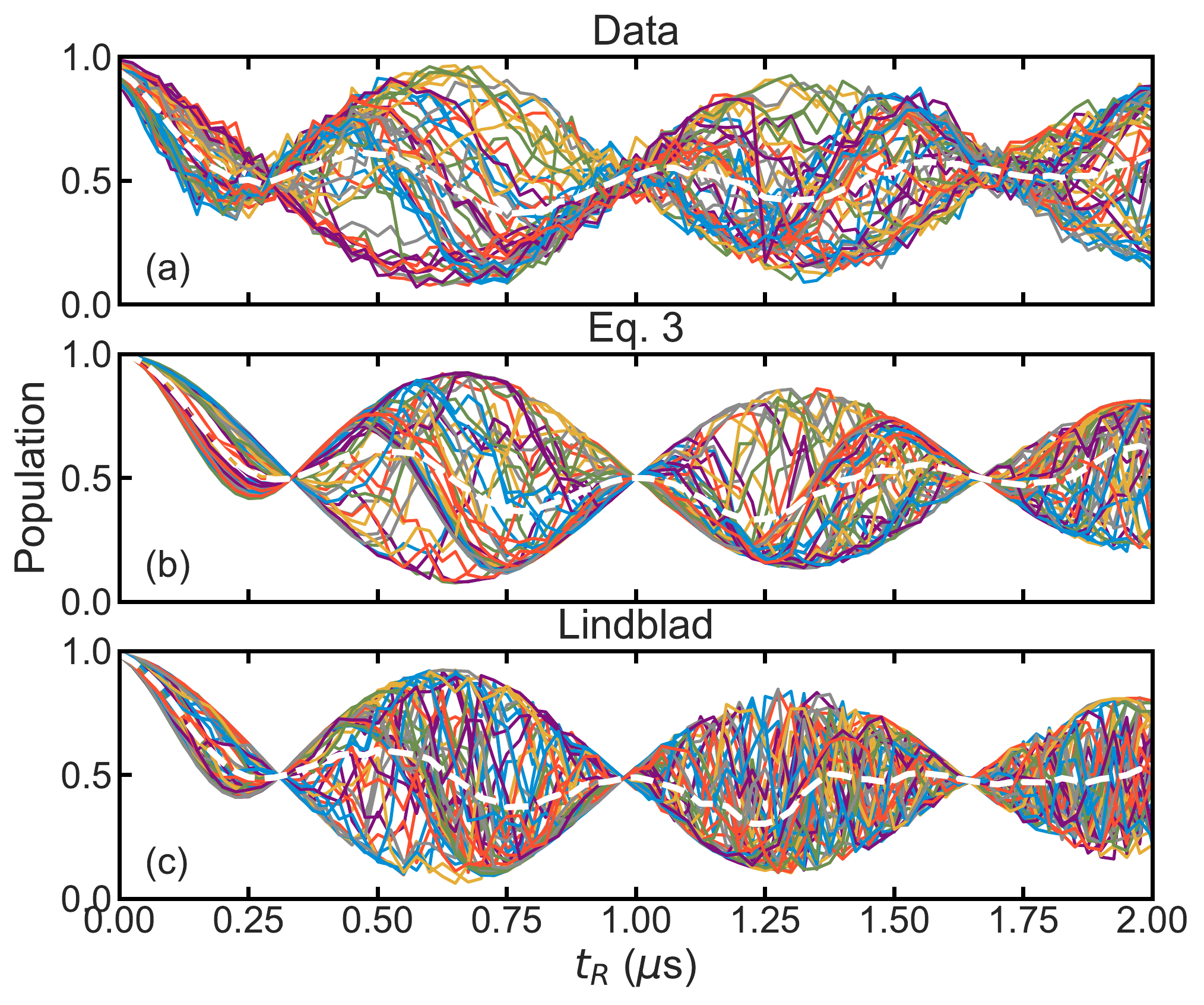}    \caption{Comparison between (a) experimental and (b,c) simulated Ramsey curves. Ramsey detuning frequency was set to $750$ kHz; white-dashed lines represent the 50-curve averages.}
   \label{fig:750kHz_comparison}
\end{figure}

Finally, equation \eqref{eq:model} can be used as an analytical approximation for low-frequency stochastic charge variations, avoiding having to solve the full Linblad master equation. To validate this, we checked the consistency of equation \eqref{eq:model} with Lindblad master equation simulations for the density matrix. 50 Ramsey curves on the 2-3 level were generated by solving the Lindblad master equation at each time step of the noise timeseries. Again, the simulations used the corrected dephasing values, and $\alpha = 2, \text{and} ~ A = 2.7\times10^{-5}e^2$/Hz for the noise parameters \cite{suppmaterial}. Comparing the experimental Ramsey curves collected on the 2-3 level with both, the 50 simulated curves generated by solving the Lindblad equation and the 50 curves generated with equation \eqref{eq:model}, the Lindblad-simulated curves qualitatively match the behavior of the experimental data as well as the curves generated with equation \eqref{eq:model} (figure \ref{fig:750kHz_comparison}).

In summary, we discussed how parity switches \cite{serniak2018hot} and charge drift can lead to ambiguous $T_2^*$ measurements, and presented a phenomenological Ramsey curve model which encapsulates the slow-varying charge fluctuations and yields measurements of $T_2^*$ which more accurately represent the total dephasing time in the higher levels of transmon qubits. Furthermore, we presented a practical approach for probing the color of the noise power spectral density by identifying unique beat patterns in the overlay of many Ramsey curves. These results should be of interest for qudit-based quantum computing platforms which aim to utilize the higher levels of transmon qubits in order to expand the computational (Hilbert) space.

See \hyperref[sec:supplementary material]{supplementary material} for measurements and simulations details. 

This work was supported by the U.S. Department of Energy (DOE), Office of Science, Basic Energy Sciences (BES) under Award DE-SC0020313 and at Lawrence Livermore National Laboratory (LLNL) under Contract No. DE-AC52-07NA27344 by DOE, Office of Science, BES, Materials Sciences and Engineering Division. We thank MIT-Lincoln Laboratories and IARPA for providing the Traveling Wave Parametric Amplifier used in this study. Experimental characterization was made possible by the National Nuclear Security Administration Advanced Simulation and Computing Beyond Moore's Law (NA-ASC-127R-16) program support of the LLNL Quantum Design and Integration Testbed (QuDIT). The numerical simulations were performed under support from DOE, Office of Science, Office of Advanced Scientific Computing Research, under project TEAM (Tough Errors Are no Match) (SCW-1683.1).(LLNL-JRNL-841934).

\clearpage
\pagebreak
\appendix*{\bf {Supplementary material: Noise-specific beats in the higher-level Ramsey curves of a transmon qubit}
\label{sec:supplementary material}}

\subsection{Multi-level Ramsey curve measurements in a 3D transmon platform}
\begin{figure}[t]
   \centering
   \includegraphics[width = .495\textwidth]{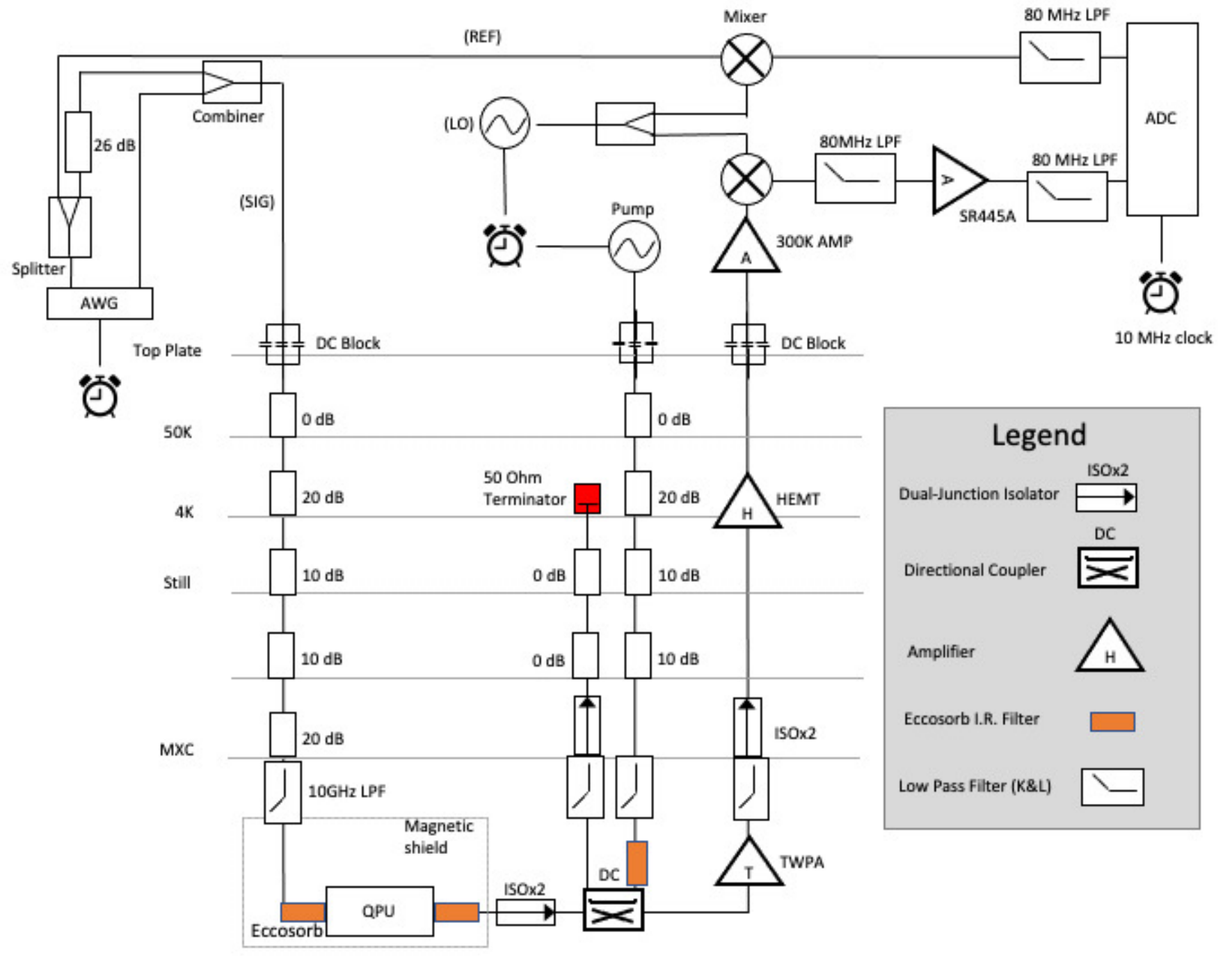} % requires the graphicx package
   \caption{Experimental setup of the multi-level 3D transmon system used to collect the Ramsey curves.}
   \label{fig:setup} 
\end{figure}

The experimental platform was a 3D aluminum on silicon superconducting transmon qubit housed in a 3D ``clamshell'' superconducting aluminum RF cavity \cite{paik2011observation}. The readout scheme was a heterodyned configuration with an intermediate frequency of 25 MHz. The amplification chain included a near quantum-limited TWPA pre-amplifier \cite{macklin2015near} at the mixing chamber of the a dilution refrigerator held below 10 mK, followed with a 4K HEMT at the 4-Kelvin stage and additional room temperature amplifiers before the 2 GHz ADC. The wiring and equipment details are summarized in figure \ref{fig:setup}. For further details and specifications on the qubit/cavity quantum characterization and readout the reader is referred to references \cite{wu2020high, martinez2020improving}. The multi-level Ramsey fringe measurement consisted of a preliminary ladder-type sequence as in reference \cite{peterer2015coherence}. First, we applied a sequence of $\{\pi_{0,1},\pi_{1,2},..,\pi_{ij}\}$ excitation pulses which excited the qubit from the ground state to the desired $j^\text{th}$ excited state where the typical two-$\pi/2$-pulse Ramsey sequence, $R_x(\pi/2)-t_R-R_x(\pi/2)$, was then applied between the $j^\text{th}$ and the $j^\text{th}+1$ level. To avoid qubit readout problems associated with the shorter $T_1$ values of the higher levels, a de-excitation sequence for ground state readout was also then applied.

\subsection{Python Simulations}
We describe the protocol, coded in python, used to simulate the Ramsey curves via equation \eqref{eq:model} along with the simulated $1/f^\alpha$ noise and the mean-square error metric used to fit the power spectral density. To account for the low frequency noise we assumed the charge drift associated with the low-frequency charge noise occurs at timescales much longer than the 1 kHz single-shot repetition rate. First, we generated a noise timeseries modeling $n(t) = q(t)/2e$ for the total duration of the acquisition. Here the $t$ is discretized by 1 ms intervals, corresponding to the 1 kHz single-shot repetition rate. Therefore, for each single-shot measurement there is a corresponding value in the noise timeseries. Each Ramsey curve consist of 101 free-evolution times sampled uniformly in $t_R\in[0,T_R]$, and at each timestep we averaged over 256 shots before incrementing to the next Ramsey timestep. The qudit Ramsey population was calculated by sampling $n(t)$ from the noise timeseries and evaluating equation \eqref{eq:model}. The evaluation sequence was carried out in identical fashion to the experimental measurement sequence described in figure \ref{fig:meas_scheme} for the total duration of the acquisition. Note that the Ramsey detuning frequency $\Omega_R$ was set experimentally, and the $T_2^*$ values were inferred from fitting to the envelope of the observed beating pattern as described in the main text. 

Before discussing the best-fit PSD procedure, we briefly discuss the scaling factor used to simulate the PSD. Since the noise timeseries amplitude was scaled to unit variance, a multiplicative scaling amplitude was used to scale the 1 Hz spectral noise amplitude ($A= S(1Hz)$) accordingly. The conversion between the dimensionless scaling amplitude $a$ and power density $A$ follows from the Fourier Transform in the PSD calculation. Let ${n}(t,\alpha)$ be the time domain representation of the $1/f^\alpha$ noise signal. For the noise simulations, we varied the timeseries noise amplitude by varying the scaling $a$, i.e., $a\times {n}$ is the charge noise signal. The PSD is given by,

\begin{equation}
S(f,\alpha) = \left|\int {a\times n(t,\alpha)} e^{-i2\pi ft} dt\right|^2.
\end{equation}
Since the type of noise in the time domain is set a priori (i.e. $1/f^\alpha$ noise ), its functional form in the frequency domain is $S(f,\alpha) =a^2 \times c_\alpha /f^\alpha$, where we explicitly defined the coefficient $c_\alpha =  \int ({n(t,\alpha)} e^{-i2\pi (1\text{Hz})t} dt)|^2$. That is, $c_\alpha$ represents the unscaled 1 Hz spectral noise amplitude which would result from a unit-variance noise timeseries. From $S(f,\alpha) = A/f^\alpha $, the scaled 1 Hz spectral noise amplitude is $A= c_\alpha\times a^2$. Note that this is expected, physically, since the power scales as the square of the amplitude. 

\begin{table*}[t!]
\begin{center}
\begin{tabular}{|l|c|c|c|c|c|c|c|c|c||c|c|}
\hline
$f_{01}$ &  $f_{12}$  & $f_{23}$& $T_{1,1} $ &  $T_{{1,2}}$& $ T_{{1,3}}$ & $T_{{2,1}}$&  $T_{{2,2}}$ & $T_{{2,3}}$ \\
\hline
4.0108 GHz & 3.8830 GHz & 3.6287 GHz & 45 $\mu s$ & 21 $\mu s$ & 22 $\mu s$ &  24 $\mu s$ &14.5$\pm4 ~\mu s$ &4.3$\pm 2.4 ~\mu s$\\ \hline
\end{tabular}
\caption{Measured values for the different parameters. In our forward simulations, we used $T_{{2,2}} = 14.5~ \mu s$ and $T_{{2,3}}= 4.3~ \mu s$, which correspond to the best-fit values. The value of charge dispersion used was $\epsilon_{23}=1.3$ MHz, which corresponds to the experimentally measured value.}
\label{tab:1}
\end{center}
\end{table*}

To find the best fit power spectral density between the experimental data and curves generated with equation \eqref{eq:model}, a power-law noise timeseries for various noise exponents $\alpha$ was generated with the algorithm described in reference  \cite{timmer1995generating}. Next, we swept the noise exponent and the scaling amplitude. The exponents varied from $0\leq\alpha\leq 3$ in steps of 0.1, and the scaling amplitude varied from $[10^{-3}$,$10^2]$, logarithmically. Continuing, we calculated the mean-squared difference, $\sum_{i=0}^n{(x_i-y_i)^2}$, where the array $x$ represents the data and array $y$ the simulated curves, for 3 parameters that comprised our best-fit metric. The first parameter was the mean-squared difference between the FFTs of the individual Ramsey curves, this step discerns high frequency spectral features of the individual curves. The second value was the mean-squared difference between the envelopes of the overlaid Ramsey curves. The envelopes of the 50-curve overlay were extracted with a simple python algorithm which extracted the maximums and minimums. This step captures the overall decay and slow varying beating features of the multi-curve overlay. Finally, the mean-squared difference of the ensemble-averaged Ramsey curves was used to capture the long-term averaged behavior. We defined our total error parameter as the equal-weighted sum of these 3 parameters, the logarithm of the sum was used since the error in the region of interest were relatively small.

\begin{figure}[t]
   \centering
   \includegraphics[width = .38\textwidth]{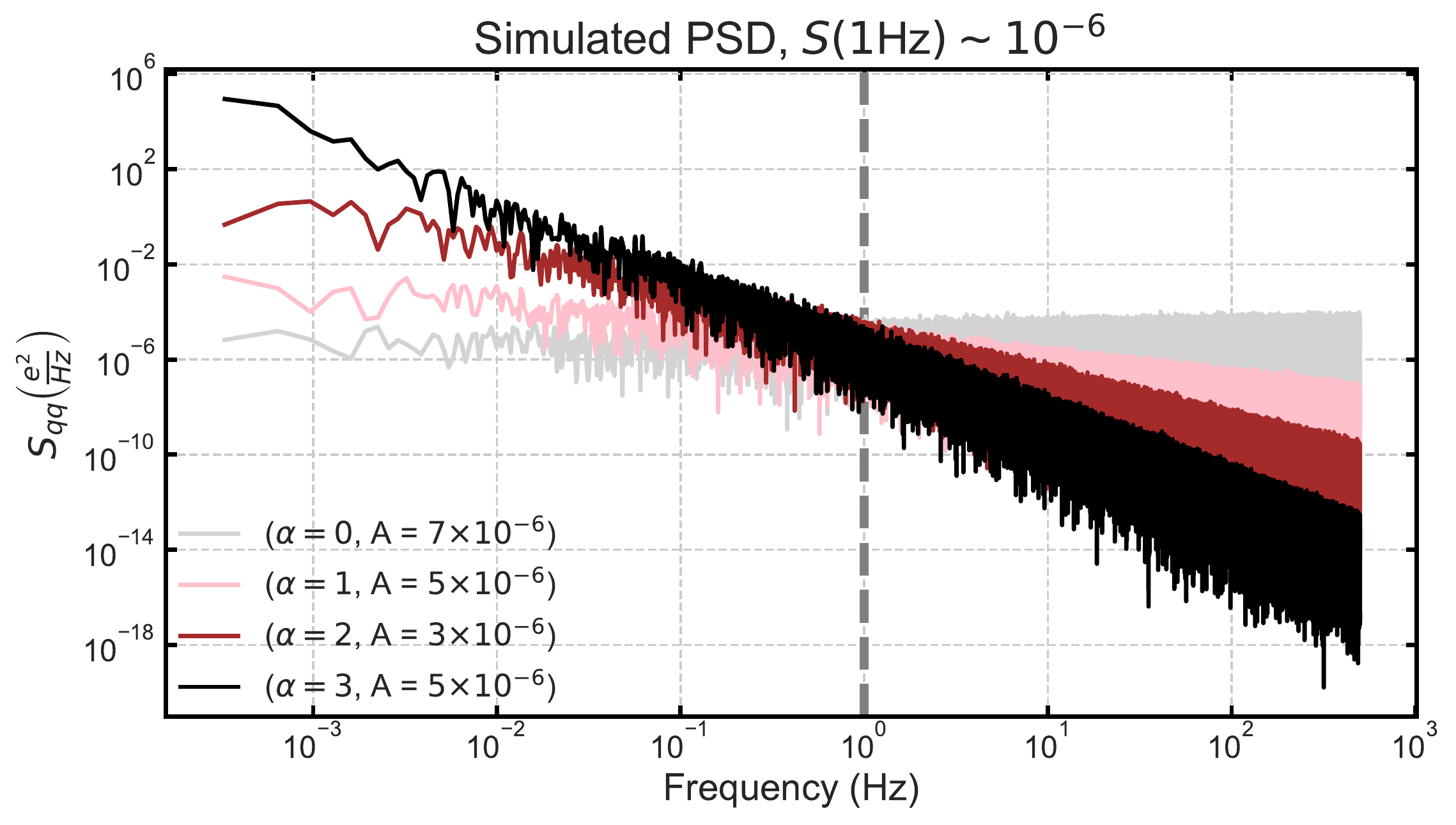} % requires the graphicx package
   \caption{Selected simulated power spectral densities for $S(1\text{Hz}) \sim 10^{-6}$, power spectrum drops off with a slope $\alpha$.}
   \label{fig:psdexample}
\end{figure}

Finally, selected simulated power spectral densities are shown in figure \ref{fig:psdexample}. The PSDs are selected so that their 1 Hz amplitude coincide at $A\sim 10^{-6}$. For low noise exponents, the high frequency noise dominates. As the noise exponent increases the higher frequency noise falls off with slope $\alpha$ and the low-frequency noise contribution becomes significant.

\subsection{Lindblad Simulations}

Here we describe the Lindblad simulations. Our model is based on the Hamiltonian $H(t) = H_{0} + H_{\rm c}(t)$, where 
\begin{equation}
H_{0} = \left[
\begin{array}{cccc}
0 & 0 & 0 & 0 \\
0 & \tilde{f}_{01} & 0 &0 \\
0 & 0 & \tilde{f}_{01} + \tilde{f}_{12}  & 0 \\
0 & 0 & 0 & \tilde{f}_{01} + \tilde{f}_{12} +\tilde{f}_{23} 
\end{array}
\right]
\end{equation}
is the system Hamiltonian and $H_{\rm c}(t)$ is the control Hamiltonian. 
Here, 
\begin{equation}
\widetilde{f}_{j,j+1} = f_{j,j+1}-\epsilon_{j,j+1}\cos\left(2\pi\frac{q}{2e}+p\pi\right),
\end{equation}
where $p\in\{0,1\}$, $f_{j,j+1}$ is the transition frequency, $\epsilon_{j,j+1}$ is the charge dispersion, $q$ is the charge over the Josephson junction, $e$ is the charge of an electron and $p$ is the parity. In principle $\frac{q}{2e}$ and $p$ can be seen as a random variables but here we assume the parity event can be averaged and treated deterministically by carrying out two simulations with $p=0$ and $p=1$ and taking their average. We take $\frac{q}{2e}$ to be a be Wiener process which we explicitly account for in our forward simulations (see below). 

Incorporating loss due to Markovian interactions with the surrounding environment leads us to a model governed by a Lindblad master equation for the density matrix $\rho$ 
\begin{equation} 
\dot{\rho} = -i\left(H\rho - \rho H\right) + \sum_{j=1}^2 \left( {\cal L}_{j} \rho {\cal L}_{j}^\dagger -
\frac{1}{2}\left( {\cal L}_{j}^\dagger{\cal L}_{j}\rho + \rho{\cal L}_{j}^\dagger{\cal L}_{j} \right) \right).
\end{equation}
Here, decay and dephasing mechanisms are modeled through the operators ${\cal L}_{1}$ and ${\cal L}_{2}$,

\begin{align}
{\cal L}_{1} &= \left[
\begin{array}{cccc}
0 & \sqrt{\gamma_{1,1}} & 0 &0 \\
0 & 0 & \sqrt{\gamma_{1,2}}  & 0 \\
0 & 0 & 0 & \sqrt{ \gamma_{1,3}} \\
0 & 0 & 0 & 0 
\end{array}
\right], 
\nonumber
\\
{\cal L}_{2} &= \left[
\begin{array}{cccc}
0 & 0 & 0 & 0 \\
0 & \sqrt{\gamma_{{2,1}}} & 0 &0 \\
0 & 0 & \sqrt{\gamma_{{2,2}}}  & 0 \\
0 & 0 & 0 & \sqrt{\gamma_{{2,3}}} 
\end{array}
\right].
\end{align}
The measured parameters used in the equations above are given in Table \ref{tab:1}. The Lindblad Ramsey curve simulations are carried out in similar sequence to simulations using equation \eqref{eq:model}, except for one crucial step.  Instead of evaluating equation \ref{eq:model} for each sampled noise value, we now solved the Lindblad equation at each timestep. The average curves of the experimental data and computed results using the Lindblad master equation and the phenomenological model are displayed in figure \ref{fig:750kHz_comparison} of the main text.

\pagebreak

\bibliography{bibliography}

\end{document}